\documentclass{JINST}

\title{A polarization-based Thomson scattering technique for burning plasmas}

\author{E. Parke\thanks{Corresponding
author.}~, V. V. Mirnov and D. J. Den Hartog\\
\llap{}Department of Physics, University of Wisconsin-Madison,\\
  Madison, WI 53706, USA\\
  E-mail: \email{eparke@wisc.edu}}

\abstract{The traditional Thomson scattering diagnostic is based on measurement of the wavelength spectrum of
scattered light, where electron temperature measurements are inferred from thermal broadening of the
scattered laser light. At sufficiently high temperatures, especially those predicted for ITER and other
burning plasmas, relativistic effects cause a change in the polarization state of the scattered photons.
The resulting depolarization of the scattered light is temperature dependent and has been proposed
elsewhere as a potential alternative to the traditional spectral decomposition technique. Following
similar work, we analytically calculate the degree of polarization for incoherent Thomson scattering.
For the first time, we obtain exact results valid for the full range of incident laser polarization states and
electron temperatures. While previous work focused only on linear polarization, we show that
circularly polarized incident light optimizes the degree of depolarization for a wide range of
temperatures relevant to burning plasmas. We discuss the feasibility of a polarization based Thomson
scattering diagnostic for ITER-like plasmas with both linearly and circularly polarized light and
compare to the traditional technique.}


\begin{document}

\section{Introduction and background}

Standard Thomson scattering diagnostics are based on wavelength spectrum measurements of scattered light from plasmas. Thermal broadening of the laser light is used to infer electron temperature for a wide range of operating conditions from a few eV to greater than 10 keV \cite{Hutchinson, Sheffield}. Polychromator systems with avalanche photodiodes have become common tools for spectrally binning the scattered light and achieving high sensitivity \cite{GApoly, HsiehAPD}. Polychromator spectral sensitivity is crucial to determining the optimal range for temperature measurements, with core and edge Thomson diagnostics relying on significantly different filter sets. For core Thomson scattering systems observing high temperature plasmas,  increased thermal broadening of the scattered light as well as variable plasma conditions can necessitate the use of a large number of spectral bins or channels.

In addition to thermal broadening of the scattered spectrum, relativistic effects at high temperature also lead to significant changes to the polarization state of scattered photons. The resulting depolarization of the scattered light can depend strongly on the electron temperature, leading other authors to propose Thomson polarization techniques as an alternative temperature measurement diagnostic \cite{Orsitto, Segre}. The proposed techniques involve up to four measurements of the polarization properties of the scattered light--the Stokes vector components--and have the potential to be simpler to implement. Expected diagnostic error bars calculated in Ref.~\cite{Orsitto} compared well with spectrally resolved Thomson predictions for fusion-grade temperatures.

Previous results however, lack a description of the depolarization of scattered light that is valid for all experimental parameters (scattering angle, temperature, and incident laser polarization). Furthermore, the work in Ref.~\cite{Segre} contains a mathematical error. The results and analysis in this paper are based on work extending the approach developed in Ref.~\cite{Segre} and presented at the IAEA Fusion Energy Conference in 2012 \cite{MirnovFEC}. For the first time, an exact analytic description of the degree of polarization has been calculated. The full derivation will be published elsewhere; here we present a brief outline of the approach and key theoretical results. We then apply these results to the diagnostic proposed in Ref.~\cite{Orsitto}.

\section{Theoretical Results}

For the typical, wavelength resolved Thomson scattering diagnostic, the scattered wave in the far-field for a single electron is Fourier transformed, and then integrated over the electron distribution function \cite{Hutchinson, Sheffield}. The resulting analytical spectrum includes both thermal broadening effects, blue-shift due to electron-headlighting, and a term commonly referred to as a "depolarization factor" which includes both relativistic headlighting effects and a reduction of scattered spectral intensity due to polarization effects. Rather than utilizing this approach, we follow the Stokes vector and Mueller matrix formalism of Ref~\cite{Segre}.

Instead of Fourier transforming the scattered field of the single electron, we express it in Stokes vector form, $\mathbf{S} = (S_{0},~S_{1},~S_{2},~S_{3})$. Here, the $S_0$ component corresponds to the total intensity of the wave and the remaining components describe the polarization properties. While the scattered light for a single electron remains completely polarized for fully polarized incident light ($S_{0}^{2} = S_{1}^{2}+S_{2}^{2}+S_{3}^{2}$), this is not the case for scattering off many electrons. From the incident and scattered electric fields, we construct a 4$\times$4 Mueller matrix describing the scattering process:  $\mathbf{S}^{(s)} = \mathbf{M} \cdot \mathbf{S}^{(i)}$. We then integrate this matrix over the electron distribution function (assumed here to be an isotropic, relativistic Maxwellian). Most of the matrix elements are zero or integrate to zero. The relevant terms of the Mueller matrix are:
\begin{eqnarray}
M_{00} &=& 1+u^2  - 2 G(\mu)(u^2 + 4 u -3) + (16/\mu^2)(1 -u)^2\nonumber \\
M_{01} &=& M_{10}=1-u^2  \nonumber \\
M_{11} &=& 1+u^2+ 2 G(\mu)(u^2 - 4 u +1)+ (12/\mu^2)(1 -u)^2\nonumber \\
M_{22} &=& 2 u  - 4 G(\mu)(u^2 - u + 1)- (12/\mu^2)(1 -u)^2\nonumber \\
M_{33} &=& 2 u  - 4 G(\mu)u(2 u -1)- (8/\mu^2)(1 -u)^2
\label{Mueller}
\end{eqnarray}
Where $u = \cos (\theta)$ represents the scattering angle dependence and $G(\mu) = K_{1}(\mu)/(\mu K_{2}(\mu))$ represents the temperature dependence, with $\mu = m_{e}c^{2} / T_{e}$. $K_{1}$ and $K_{2}$ are modified Bessel functions of the second kind.

The work in Ref~\cite{Segre} contains an error in the power of the term $(1 - \beta_{s})$ originating in the Li\'{e}nard-Wiechert expression for the scattered field. While there is contention in the literature over the power of this term for an infinite scattering volume, for a finite volume the appropriate term is unambiguously $(1 - \beta_{s})^5$ rather than $(1 - \beta_{s})^6$. At fusion-grade temperatures this can introduce non-negligible error. The results above were calculated with the correct power.

For incident light of arbitrary intensity and polarization, the scattered light can be expressed in terms of the Mueller matrix elements: for example, $S^{(s)}_{0} = M_{00} \cdot S^{(i)}_{0} + M_{01} \cdot S^{(i)}_{1}$. Due to the nature of the electron distribution function, photons of many polarization states will be present in the scattered light. This is described by the degree of polarization, $P = \sqrt{S_{1}^{2}+S_{2}^{2}+S_{3}^{2}} / S_{0}$, and the degree of depolarization, $D = 1 - P$.

\subsection{Diagnostic implications}

From the theoretical results above, a few constraints on diagnostic design can be determined. Both scattering angle and incident laser polarization are important parameters for Thomson scattering diagnostics. Most diagnostics operate with scattering angles near $\theta = \pi/2$ but covering a wide range, while the LIDAR Thomson system proposed for ITER would operate near $\theta = \pi$ \cite{ITERlidar}. Universal (to our knowledge) use of linearly polarized incident light with the electric field aligned perpendicular to the scattering plane (parallel to the toroidal axis) originates in the simplicity offered to scattering spectrum calculation as well as optimization of scattered intensity for cold electrons.

\begin{figure}
\centering
\includegraphics[scale = 0.5]{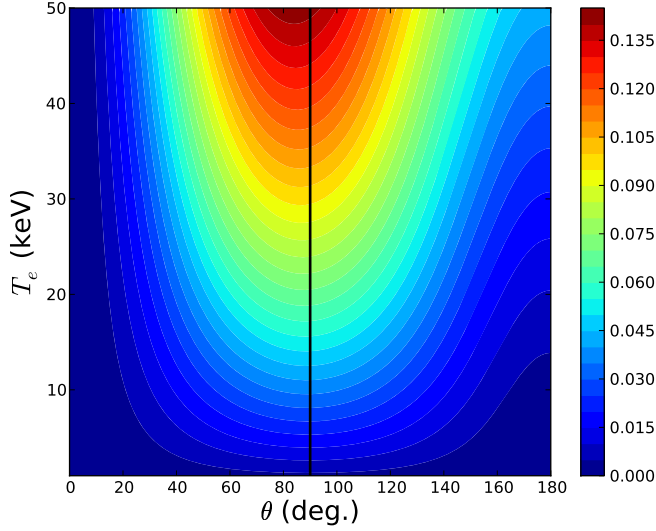}
\caption{Degree of depolarization for linearly polarized incident light with $\psi = 0$.}
\label{LinearDepol}
\end{figure}

\begin{figure}
\centering
\includegraphics[scale = 0.5]{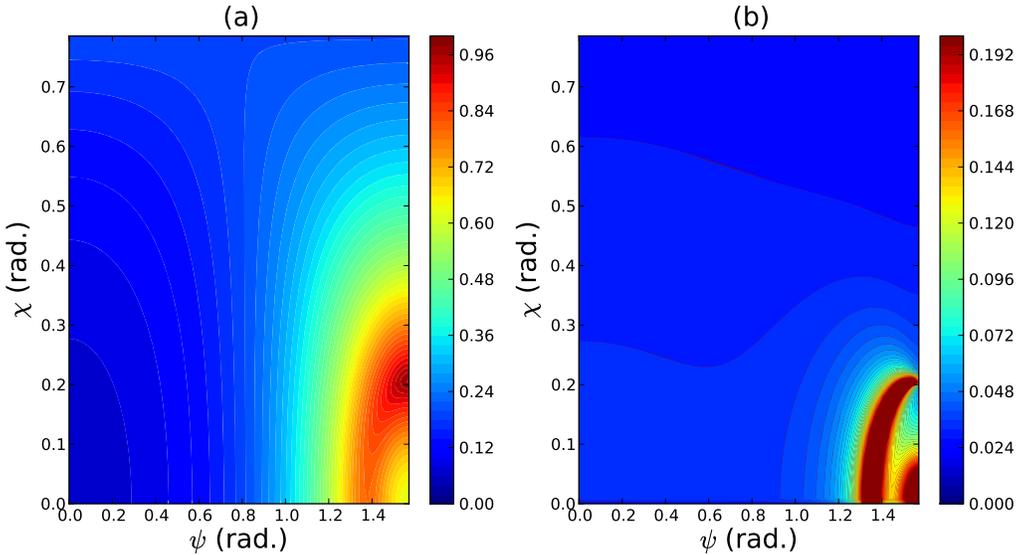}
\caption{(a) Depolarization and (b) predicted error in temperature measurement for 20 keV plasmas at $\pi/2$ scattering angle with varying incident laser polarization (ellipticity $\chi$ and orientation angle $\psi$).}
\label{DepolWithErr}
\end{figure}

Figure~\ref{LinearDepol} shows the degree of depolarization across the full range of scattering angles and fusion relevant temperatures for linearly polarized light. Depolarization is strongest near perpendicular scattering, although not exactly at $\theta = \pi/2$. The exact angle of maximum depolarization is temperature dependent, and these results are consistent with the findings of Ref.~\cite{Segre} showing maximum depolarization deviating slightly from $\pi/2$. However, this is only a local maximum due to the choice of linearly polarized light--the true maximum occurs for elliptically polarized light, see Figure~\ref{DepolWithErr}(a).

Far away from $\pi/2$ scattering, the depolarization drops off rapidly. For both forward and backward scattering, the degree of depolarization is no more than a few percent for expected reactor temperatures. A polarization-based Thomson scattering technique would be highly unsuitable for diagnostics like the proposed ITER LIDAR system, while traditional Thomson scattering diagnostics near $\theta = \pi/2$ would offer greater depolarization with stronger temperature dependence.

Evaluating the effectiveness of different laser polarization parameters is slightly more complicated. Some configurations offer strong degrees of depolarization, but weak scattered intensity. In Fig.~\ref{DepolWithErr}(b), the predicted error in the temperature measurement is plotted against both the orientation angle of the laser ($\psi = 0$ for toroidal alignment) and the ellipticity ($\chi = 0$ for linear, $\chi = \pi/4$ for circular). This includes only Poisson statistics for the scattered photons and neglects background light; background signal is accounted for in the diagnostic simulations of Section 3. The region near $\psi = \pi/2$ exemplifies the high depolarization, low scattered intensity trade off: large measurement errors limit diagnostic viability and the features are highly sensitive to electron temperature, making this section of parameter space one to be avoided.

The minimum error corresponds to circular polarization. Thomson scattering literature focuses almost exlusively on linear polarization, and even Ref.~\cite{Segre} is restricted in analysis to linearly polarized light. These results highlight the versatility of this approach and the need to consider all incident polarizations. However, it should be noted that, for the purposes of a Thomson polarization diagnostic, linearly polarized light closely aligned to the toroidal axis can achieve error bars competetive with circularly polarized light.

\section{Diagnostic Design and Simulation}

\begin{figure}
\centering
\includegraphics[scale = 0.35]{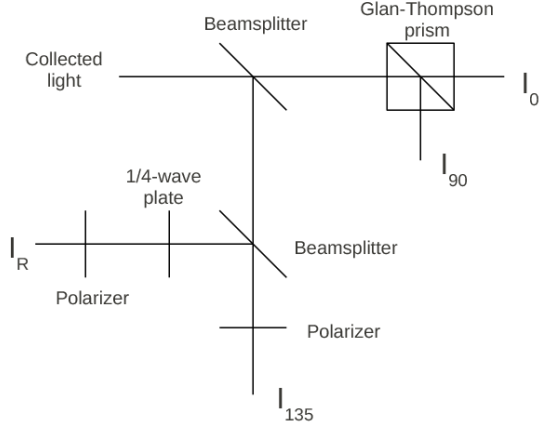}
\caption{Polarization measurement design adapted from Ref.~\cite{Orsitto}, with modifications for elliptically polarized incident light.}
\label{DetectorDesign}
\end{figure}

Using the insight developed above, we further explore the viability of a polarization-based Thomson scattering diagnostic by applying our theoretical results to the proposed design in Ref.~\cite{Orsitto}. The original design was intended for linearly polarized incident light, so the diagram in Figure~\ref{DetectorDesign} has been modified slightly to enable measurements with circularly polarized light. The four Stokes components are related to six measureable intensities, although only four of the measureable quantities are independent. For this proposal, we choose three of the intensity measurements from the original design: without phase retardation, the intensity is measured after polarization selection at angles of $0^{\circ}$, $90^{\circ}$, and $135^{\circ}$ relative to an electric field vector aligned to the toroidal axis. The only modification to the design requires removing the half-wave plate and second Glan-Thompson prism, replacing them with the beamsplitter, quarter-wave plate, and appropriate polarizers in the lower right of the diagram. This would lead to a slightly more complicated mounting arangement, but allows measurement of the right-hand circular component for elliptically polarized incident light. The Stokes vectors are calculated from the measured intensities using the following relations:
\begin{eqnarray}
S_{0} &=& I_{0^{\circ}} + I_{90^{\circ}} \nonumber \\
S_{1} &=& I_{0^{\circ}} - I_{90^{\circ}} \nonumber \\
S_{2} &=& S_0 - 2 I_{135^{\circ}} \nonumber \\
S_{3} &=& S_0 - 2 I_{R}
\label{StokesExp}
\end{eqnarray}

The simluated diagnostic utilizes a laser capable of producing 5 J pulses at 1064 nm wavelength with an integration time of 50 ns. The scattering angle is $\theta = \pi/2$ from a core location with variable electron temperature $T_{e}(0)$. We treat the background light as entirely bremstrahhlung, ignoring line radiation. To model the bremstrahhlung, the electron temperature and density profiles are assumed to be parabolic, $T_{e}(r) = T_{e}(0) \cdot (1-r^2)$, with core electron density of 1$\cdot$10$^{20}$ cm$^{-3}$. The background light is integrated along a 4 m line of sight and over wavelengths in the range 200-2000 nm to accomodate the full width of the scattered spectrum. These parameters are chosen for similarity with ITER.

The simulated diagnostic error bars are calculated for core temperatures ranging from 10 keV to 50 keV. Three cases are compared: $\chi = \pi/4$, ($\psi$,$\chi$) = (0,0), and ($\psi$,$\chi$) = ($\pi/6$,0). The circularly polarized light utilizes the full, 4-component polarimeter shown, while the linear cases utilize reduced forms: measurement of only $I_{0^{\circ}}$, $I_{90^{\circ}}$, and $I_{135^{\circ}}$ for the ($\psi$,$\chi$) = ($\pi/6$,0) case and further reduction to only $I_{0^{\circ}}$ and $I_{90^{\circ}}$ for the ($\psi$,$\chi$) = (0,0) case. The reduced forms of the polarimeter benefit from improved scattered signal amplitude on the measured channels.

\begin{figure}
\centering
\includegraphics[scale = 0.5]{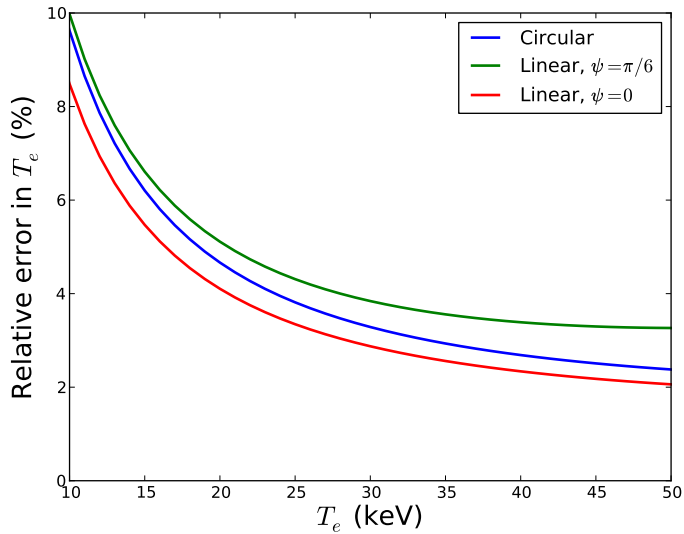}
\caption{Predicted error in temperature measurement for three implementations of the polarization diagnostic: circular, linear with $\psi = \pi/6$, and linear with $\psi = 0$.}
\label{DiagnosticErr}
\end{figure}

The results are shown in Figure~\ref{DiagnosticErr}. Above 10 keV, all cases achieve error bars of less than 10\%, and less than 5\% above 20 keV, making them competitive with standard Thomson scattering diagnostics. The circular polarization case offers the best performance for the full, 4-component polarimeter, and even bests the reduced 3-component form. However, the 2-component polarimeter achieves the best results across the full range of temperatures, with error bars approaching 2\% at 50 keV. The flatness of the curves above 20 keV indicates that these diagnostics would be robust over a wide range of fusion-relevant temperatures. However, line radiation will lead to some degradation of performance.

\section{Advantages and implementation challenges}

A polarization-based diagnostic offers several advantages over the polychromator design common now. With a maximum of 4 channels per radial position, the polarization diagnostic could translate into cost savings in detector hardware and digitizer channels. The need for fewer measurement channels would be less operationally demanding, while fewer optics make for a simpler, more robust diagnostic. The 2-component form of the polarimeter maximizes these advantages.

Unfortunately, current technology makes a polarization-based diagnostic challenging. The collection optics and fibers for transmitting the scattered light to the polarization meter must conserve the polarization state of the scattered light. Optics capable of this are available, but expensive. The polarization sensitivity of the avalanche photodiodes (or other detectors) must also be taken into account.

More significantly, while the original design in Ref.~\cite{Orsitto} makes optional use of a half-wave plate to simplify the arrangement of detectors for the $I_{-45^{\circ}}$ and $I_{+45^{\circ}}$ components, the full, 4-component polarization meter suggested here requires the use of a quarter-wave plate to measure the $I_{R}$ component of the scattered light. While the scattered spectrum covers wavelengths spanning a range of hundreds of nanometers, currently available waveplates do not have a uniform response over such a wide range. Integration over the full range of wavelengths is necessary for operation of such a diagnostic--both to obtain sufficient signal intensity and in order for the Mueller matrix representation to be valid. The non-uniform response of the wave plate makes the polarization meter unfeasible for circularly polarized light.

While the 3- and 2-component polarimeters for linear incident light do not suffer from the non-uniform responses of the wave plates, the width of the scattering spectrum at high temperatures still presents technical challenges. Nd:YAG lasers at 1064 nm are near the cut-off in responsivity for Si avalanche photodiodes. The red-shifted part of the spectrum falls outside the measurement capabilities of detector hardware common in Thomson scattering systems optimized for measuring blue-shift. Even the available Glan-Thompson prisms for near-IR wavelengths do not have uniform response over thousands of nanometers. Frequency-doubled Nd:YAG lasers at 532 nm would have narrower scattered spectra, but the spectrum widths at such temperatures would still be appreciable and extend deep into the UV.

\acknowledgments

This work supported by the U.S Department of Energy.

\end{document}